\documentclass[usenatbib]{mn2e}
\usepackage{natbib}
\usepackage{amssymb}
\usepackage{epsfig}
\usepackage{rotating} 
\usepackage{aas_macros}

\newcommand{\Lx}{\ensuremath{L_{\mathrm{X}}}}
\newcommand{\Yx}{\ensuremath{Y_{\mathrm{X}}}}

\newcommand{\Msol}{\ensuremath{\mathrm{M_{\odot}}}}

\newcommand{\rf}{\ensuremath{R_{\mathrm{500}}}}
\newcommand{\rn}[1]{\ensuremath{R_{\mathrm{#1}}}}

\newcommand{\Zsol}{\ensuremath{\mathrm{Z_{\odot}}}}
\newcommand{\fgas}{\ensuremath{f_{\mathrm{gas}}}}

\newcommand{\OM}{\ensuremath{\Omega_{\mathrm{M}}}}

\newcommand{\egc}{{\it e.g.}}  
\newcommand{\etal}{et al.\ }

\newcommand{\Chandra}{\emph{Chandra}}

\newcommand{\XMM}{\emph{XMM-Newton}}

\newcommand{\chisq}{\ensuremath{\chi^2}}

\newcommand{\gta}{\,\rlap{\raise 0.4ex\hbox{$>$}}{\lower 0.6ex\hbox{$\sim$}}\,}  
\newcommand{\lta}{\,\rlap{\raise 0.4ex\hbox{$<$}}{\lower 0.6ex\hbox{$\sim$}}\,}  

\newcommand{\nm}{\mbox{\ensuremath{\mathrm{~\nm}}}}

\newcommand{\cm}{\mbox{\ensuremath{\mathrm{~cm}}}}

\newcommand{\km}{\mbox{\ensuremath{\mathrm{~km}}}}

\newcommand{\kpc}{\mbox{\ensuremath{\mathrm{~kpc}}}}
\newcommand{\Mpc}{\mbox{\ensuremath{\mathrm{~Mpc}}}}
\newcommand{\s}{\mbox{\ensuremath{\mathrm{~s}}}}
\newcommand{\ks}{\mbox{\ensuremath{\mathrm{~ks}}}}

\newcommand{\yr}{\mbox{\ensuremath{\mathrm{~yr}}}}

\newcommand{\keV}{\mbox{\ensuremath{\mathrm{~keV}}}}
\newcommand{\erg}{\mbox{\ensuremath{\mathrm{~erg}}}}

\newcommand{\MHz}{\mbox{\ensuremath{\mathrm{~MHz}}}}

\newcommand{\arcm}{\ensuremath{\mathrm{^\prime}}}
\newcommand{\arcs}{\arcm\hskip -0.1em\arcm}

\newcommand{\K}{\mbox{\ensuremath{\mathrm{~K}}}}

\newcommand{\pcc}{\ensuremath{\mathrm{\cm^{-3}}}}

\newcommand{\pcmsq}{\mbox{\ensuremath{\mathrm{~cm^{-2}}}}}
\newcommand{\pMpc}{\ensuremath{\mathrm{\Mpc^{-1}}}}
\newcommand{\ps}{\ensuremath{\mathrm{\s^{-1}}}}

\newcommand{\ergps}{\ensuremath{\mathrm{\erg \ps}}}

\newcommand{\kmpspMpc}{\ensuremath{\mathrm{\km \ps \pMpc\,}}}

\newcommand{\YM}{\mbox{\ensuremath{\mathrm{Y_{X}-M}}}}

\newcommand{\LM}{\mbox{\ensuremath{\mathrm{L_X-M}}}}

\newcommand{\MT}{\mbox{\ensuremath{\mathrm{M-kT}}}}

\newcommand{\LCDM}{$\Lambda$CDM~}

\usepackage{url}

\newcommand{\jjj}{XLSSC 029}
\newcommand{\jjjj}{XLSSJ022403.9-041328}

\begin{document}


\title[\XMM\ and \Chandra\ observations of \jjjj]{Testing the galaxy cluster mass-observable relations
at $z=1$ with \XMM\ and \Chandra\ observations of \jjjj\thanks{Based on
observations obtained with \XMM, an ESA science mission with instruments
and contributions directly funded by ESA Member States and NASA.}.}
\author[B.J. Maughan \etal]{
\parbox[h]{\textwidth}{
B. J. Maughan,$^{1,2}$\thanks{E-mail: ben.maughan@bristol.ac.uk}\thanks{Chandra Fellow}
    L. R. Jones,$^3$ M. Pierre,$^4$ S. Andreon,$^5$ M.
Birkinshaw,$^2$ M. N. Bremer,$^2$ F. Pacaud,$^6$ T. J. Ponman,$^3$
I. Valtchanov,$^7$ and J. Willis$^8$
}
\vspace*{12pt} \\
\parbox[h]{\textwidth}{
  $^1$Department of Physics, University of Bristol, Tyndall Ave, Bristol BS8 1TL, UK.\\
  $^2$Harvard-Smithsonian Center for Astrophysics, 60 Garden St, Cambridge, MA 02140, USA.\\
  $^3$School of Physics and Astronomy, The University of Birmingham, Edgbaston, Birmingham B15 2TT, UK.\\
  $^4$Service d'Astrophysique, CEA Saclay, 91191 Gif sur Yvette, France.\\
  $^5$INAF-Osservatorio Astronomico di Brera, Via Brera, 28, 20121 Milano, Italy.\\
  $^6$Argelander Institute for Astronomy, Bonn University, Auf dem Hügel 71, 53121 Bonn, Germany.\\
  $^7$ESA European Space Astronomy Centre, P.O. Box 78, 28691 Villanueva de la Ca\~nada, Madrid, Spain.\\
  $^8$Department of Physics \& Astronomy, University of Victoria, Elliot 
Building, 3800 Finnerty Road, Victoria, BC, V8P 1A1, Canada.\\
}}

\maketitle

\begin{abstract}
We present an analysis of deep \XMM\ and \Chandra\ observations of the
$z=1.05$ galaxy cluster \jjjj\ (hereafter \jjj), detected in the \XMM\ large scale structure
survey. Density and temperature profiles of the X-ray emitting gas were
used to perform a hydrostatic mass analysis of the system. This allowed us
to measure the total mass and gas fraction in the cluster and define
overdensity radii \rf\ and \rn{2500}. The global properties of \jjj\ were
measured within these radii and compared with those of the local
population. The gas mass fraction was found to be consistent with local
clusters. The mean metal abundance was $0.18^{+0.17}_{-0.15}\Zsol$, with the
cluster core regions excluded, consistent with the predicted and observed
evolution. The properties of \jjj\ were then used to investigate the
position of the cluster on the \MT, \YM, and \LM\ scaling relations. In all
cases the observed properties of \jjj\ agreed well with the simple
self-similar evolution of the scaling relations. This is the first test of
the evolution of these relations at $z>1$ and supports the use
of the scaling relations in cosmological studies with distant galaxy
clusters.
\end{abstract}


\section{Introduction} \label{s.intro}
The mass function of galaxy clusters is predicted by theoretical models of
the formation of structure from the density fluctuations in the early
universe. The shape and evolution of the predicted mass function are
strongly dependent on the details of the input models, and so comparison of
predicted and observed mass functions at different redshifts can place
tight constraints on the values of interesting cosmological parameters
\citep[e.g.][]{rei02,vik03,hen04}. However, in order to perform these
tests, a key observational challenge must be met; galaxy cluster masses
must be determined from their observable properties. This is an area of
active research, and many approaches have been investigated using different
observables, including the properties of the hot ionised intra-cluster
medium (ICM) as measured from its X-ray emission \citep{sar86,ros02} or the
Sunyaev-Zel'dovich effect \citep{bir99}; the distribution of properties of member
galaxies such as richness and velocity dispersions and the strong and weak
gravitational lensing effects of cluster gravitational potentials on
background galaxies \citep[e.g.][]{smi01b,dah06a}. Numerical simulations of
individual galaxy clusters and large cosmological volumes are also an
invaluable tool for studying clusters and testing these different
observational techniques \citep[e.g.][]{evr96,nag07}.

X-ray observations of clusters are useful because the ICM is extremely
X-ray luminous, allowing the detection and study of galaxy clusters out to
$z\ga1$. Under the assumption that the ICM is in hydrostatic equilibrium
with the cluster's gravitational potential \citep[a reasonable assumption
for clusters that are not actively merging;][]{poo06} then radial profiles
of the gas density and temperature measured from X-ray data can be used to
determine the total cluster mass within some radius. The radius used is
typically chosen to enclose an overdensity $\Delta$ with respect to the
critical density at the cluster's redshift, with $\Delta=200$ approximating
to the virial radius, and $\Delta=500$ the maximum radius detectable in
typical observations. The hydrostatic mass estimation described above
demands high quality X-ray data with which to measure the required
profiles. Such data are becoming more commonplace in the \Chandra\ and
\XMM\ era, but are still far from the norm. In general, and
particularly for distant clusters (which provide the most information
for measuring cosmological parameters), X-ray data permit the
measurement of simple global properties such as a single gas
temperature and luminosity. Such properties can still be useful for
estimating cluster masses, as power law scaling relations exist
between global properties and cluster mass \citep{kai86,mar98a,fin01}. These scaling
relations are predicted by simple self-similar models of
clusters. While the observed relations can differ from the
self-similar predictions, the slope, normalisation and evolution of
the relations can be measured using high quality data, and then
applied to give mass estimates for poorer quality data. These
techniques have allowed studies using the temperature and luminosity
functions of galaxy clusters as cosmological probes, with scaling
relations used to convert between the observables and masses for
comparison with theoretical mass functions. 

The scaling relations are, however, an imperfect tool for converting
observables to masses. Radiative cooling and cluster mergers cause
significant scatter in the relations between X-ray luminosity (\Lx), temperature (kT) and mass, limiting the
accuracy of derived masses. Much of the scatter derives from the dense core
regions of clusters where radiative cooling can be extremely efficient, and
the effects of mergers are most pronounced. Excluding these regions can
significantly reduce the resulting scatter
\citep{mar98a,oha06,mau07b}. Recent work by \citet{kra06a} has shown that
the parameter \Yx, the product of gas temperature and gas mass has an extremely
low scatter with total mass, providing the potential for reliable cluster mass
estimates from simple observables. 

The mass-observable scaling relations are reasonably well-calibrated
for relaxed clusters in the local universe ($z<0.15$)
\citep{arn05,vik06a}, but at higher redshifts, the relations are less
well measured. This is due to the long observations required to obtain
sufficiently deep X-ray data to allow hydrostatic mass estimates
against which to plot the simple observables. Recent work, based on
small samples of clusters with the best available data has enabled the
measurement of the \MT\ relation at $z\la0.7$ \citep{kot05,kot06}. The
relations were found to have the same slope as their local
counterparts, with the evolution of the normalisation well described
by the self-similar model. Furthermore, using some of the same data,
\citet{mau07b} found that the \YM\ relation at $z\sim0.6$ is
consistent with self-similar evolution of the local relation. Beyond
$z\sim0.7$, X-ray hydrostatic mass estimates have been made for only
one cluster CLJ1226.9+3332, although that mass estimate may be biased
by merging activity in the cluster
\citep{mau07a}. The evolution of the mass-observable relations to such high
redshifts have thus not been well studied, although more and more clusters
are being detected at these distances
\citep[e.g.][]{ros04a,mul05,sta06a,bre06,pie06short} providing a powerful resource for
cosmological studies, given well-measured scaling relations. In
this paper we make an important step forward in this endeavor by measuring
the first X-ray hydrostatic mass for a cluster at $z>1$ and
investigating its position on the various mass-observable scaling relations.

Galaxy cluster \jjjj\ (hereafter \jjj) was detected as an extended X-ray source in 
the \XMM\ Large Scale Structure (XMM-LSS) survey \citep{and05,pie06short,pac07}. The XMM-LSS survey is described in \citet{pie04a}, and some results from the survey are presented in \citet{val04,wil05,pie06short,pac07}. NIR imaging confirmed an overdensity of faint galaxies
coincident with the X-ray source \jjj, and spectroscopic follow-up confirmed the
redshift of the cluster to be $z=1.05$. The cluster has since been the target of deep
\Chandra\ ACIS-S and \XMM\ follow-up observations, which are the subject of
this paper. In the following sections we describe the reduction of the
X-ray data, present the results of our mass analysis, and investigate the
location of the cluster on the different mass-observable relations. A \LCDM
cosmology of $H_0=70\kmpspMpc)\equiv 1$, and $\OM=0.3$
($\Omega_\Lambda=0.7$) is adopted throughout and all errors are quoted at
the $68\%$ level. In this cosmology, at $z=1.05$, 1 arcsecond corresponds to $8.1\kpc$.

\section{X-ray Data Reduction}
The reduction and analysis methods used were the same as those
presented in \citet{mau07a} for both the \Chandra\ and \XMM\ data, but
the most important details are repeated here. The standard data
reduction procedures were followed, using the most recent calibration
products available as of March 2007. Lightcurves were produced for the
observations, and were cleaned to remove periods of high
background. The \Chandra\ observation of \jjj\ was taken in six
consecutive exposures with cleaned exposure times of between $10\ks$ and
$30\ks$. The total \Chandra\ exposure time was $127\ks$, yielding a total of $\sim1300$ net source counts in the $0.3-5\keV$ band. The \XMM\ observations
had a cleaned exposure time of $70\ks$ (pn) and $88\ks$ (MOS) giving a total of $\sim3400$ net source counts in the $0.3-5\keV$ band. The observations used are summarised in Table \ref{t.obs}.

\begin{table*}
\centering
\begin{tabular}{ccccccccccc} \hline 
Date & Detector & Obs ID & Total Exposure (ks) & Good Time (ks) \\ \hline
2005-01-01 & \XMM\ pn & 0210490101 & 86 & 70 \\
2005-01-01 & \XMM\ MOS1 & 0210490101 & 106 & 87 \\
2005-01-01 & \XMM\ MOS2 & 0210490101 & 106 & 88 \\
2005-09-13 & \Chandra\ ACIS-S & 6390 & 12 & 10 \\
2005-10-12 & \Chandra\ ACIS-S & 6394 & 18 & 18 \\
2005-10-12 & \Chandra\ ACIS-S & 7182 & 23 & 23 \\
2005-10-14 & \Chandra\ ACIS-S & 7184 & 23 & 23 \\
2005-10-15 & \Chandra\ ACIS-S & 7183 & 20 & 20 \\
2005-11-21 & \Chandra\ ACIS-S & 7185 & 33 & 33 \\
\hline
\end{tabular}
\caption{\label{t.obs}Summary of the \XMM\ and \Chandra\ observations of \jjj.}
\end{table*}

\subsection{Background preparation}\label{sec:backgr-prep}
For the analysis of both \Chandra\ and \XMM\ data, background
estimates were derived from blank-sky datasets, using the same detector
regions as the source emission being considered. Important differences
exist between the blank-sky data and the observations of a given
source, due primarily to differences in the levels of the particle-induced background components (which can vary significantly with time
and dominate at energies $\ga2\keV$)
and differences in the soft Galactic foreground emission (which can
vary significantly with position on the sky and dominates at energies
$\la2\keV$). These differences must be corrected for when using the
blank-sky data. 

In the case of \jjj, due to the target's high redshift, the source
observations contained large detector regions that were free from
source emission. Spectra were extracted from these ``local'' regions
and compared with the blank-sky spectra from the same detector
regions.

\begin{figure*}
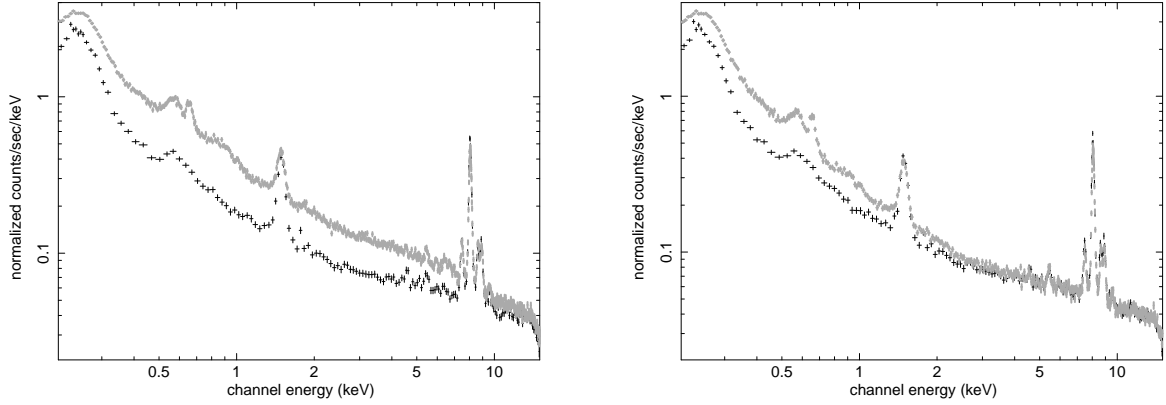

\begin{center}
\scalebox{0.3}{\includegraphics*[angle=270]{pnppsbgspec.ps}}
\hspace{1cm}
\scalebox{0.3}{\includegraphics*[angle=270]{xlss29pnbgspec.ps}}
\caption{\label{f.bgspec}Comparison of the local (black points) and blank-sky
(grey points) background spectra measured with the \XMM\ pn
camera. The two plots show the effect of filtering the blank-sky data
for background flares in the high-energy band alone (left plot) and of
additional filtering in the $2-5\keV$ band (right plot).}
\end{center}
\end{figure*}

For the \XMM\ data, the blank-sky files of \cite{car07} were
used. Periods of high particle background were removed by filtering
lightcurves produced in the $10-12\keV$ (MOS) and $12-14\keV$ (pn)
energy bands with the same count rate limits that were applied to the
source data when they were cleaned. The blank-sky spectra were then
normalised to match the local background count rates in these high
energy bands and compared with the local background
spectra. Initially, a very poor agreement was found, and additional
flare filtering at lower energies ($2-5\keV$, using iterative
$3\sigma$ clipping of the lightcurve) was required for the blank-sky
files before there was a good agreement between the local and
blank-sky background spectra. This is illustrated for the pn camera in
Fig. \ref{f.bgspec}, and very similar results were found for both MOS
cameras. After this additional filtering, the ratios of the entire
field local to blank-sky count rates in the high energy bands were
$1.01, 1.10$ and $1.14$ for pn, MOS1 and MOS2 respectively, indicating that
the particle background levels are similar in the target and blank-sky
fields. After scaling by these factors, the \XMM\ local and blank-sky
spectra agreed well except below $\sim2\keV$ where there was a significant
decrement in all local background spectra compared to the blank-sky
spectra. This is due to significantly below average soft Galactic
foreground emission in the direction of \jjj\ and is corrected for in
subsequent spectral and image analysis.

For the \Chandra\ observations, the standard blank-sky background
files were
used\footnote{\url{http://cxc.harvard.edu/contrib/maxim/acisbg/}}. The
blank-sky spectra were then normalised to the local background spectra
in the $9.5-12\keV$ band to account for variations in the particle
background and compared. A good agreement was found in all of the
Chandra observations, with no additional filtering
required. Consistent with the \XMM\ background spectra, there was a
significant decrement at $<2\keV$ in all local background spectra due
to the low soft Galactic foreground emission in the direction of \jjj.

\section{Image Analysis}\label{s.img}
\begin{figure*}
\begin{center}
\scalebox{0.8}{\includegraphics*[angle=0]{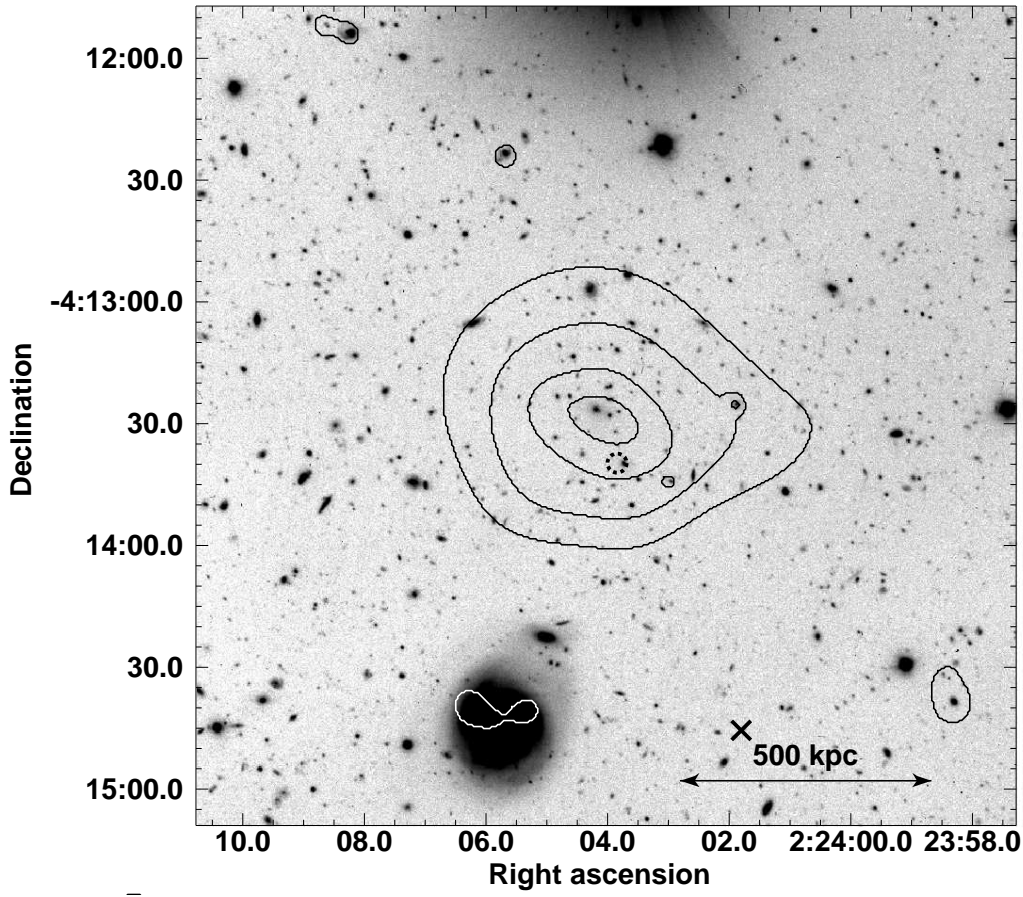}} \\
\caption{\label{f.overlay} Contours of the adaptively smoothed image
produced from the \Chandra\ data are overlaid on a VLT I-band image
of \jjj. The dotted circle shows the position of the $325\MHz$ radio
source detected by \citet{coh03} with no apparent X-ray
counterpart. The cross near the bottom right indicates the location of
a faint X-ray point source detected in the \XMM\ data that was not
detected by \Chandra.}
\end{center}
\end{figure*}

Images were produced for each dataset in the $0.3-2.0\keV$ (observed
frame) energy band, and the \Chandra\ and \XMM\ data were co-added to
make combined images for each satellite. These images were then
vignetting corrected (using an exposure map generated for $1.5\keV$
photons) and adaptively smoothed \citep{ebe06a}. Contours of the
smoothed
\Chandra\ data are plotted in Fig \ref{f.overlay}, overlaid on an image produced from a $2400\s$ FORS2 VLT I band exposure 
taken in 2005 August (proposal ID 075.A-0175, PI Andreon) with 0.7 arcsecond 
seeing. The contour levels were defined so that the
emission bounded by adjacent contours was detected at a significance
of $3\sigma$ above that bounded by the surrounding contour
\citep[see][for details]{mau07a}. The X-ray morphology is slightly elliptical, but
appears reasonably relaxed, with no indication of substructure in the
images. The X-ray morphology in the \XMM\ image was consistent, with
stronger contamination from point sources due to the larger point spread function (PSF).
The centroid of the X-ray emission (determined from the
\Chandra\ data alone) is located at $\alpha[2000.0]=02^{\rm h}24^{\rm
m}4.0^{\rm s}$ $\delta[2000.0]=-04^{\circ}13\arcm 28.9\arcs$.

Wavelet decomposition was used to detect point sources in the images \citep{vik98b},
and detected point sources were excluded from all analysis of the
extended emission. There was no detection of point-like emission from
the central galaxy. The X-ray point sources are apparent in
Fig. \ref{f.overlay} which also shows the position of a faint source
detected by \XMM\ but not by \Chandra. Finally, a low frequency radio
source was detected at $325\MHz$ (J0224.0-0413 at $\alpha[2000.0]=02^{\rm h}24^{\rm m}3.82^{\rm s}$ $\delta[2000.0]=-04^{\circ}13\arcm 38.8\arcs$), close to the core of \jjj\ by
\cite{coh03}. No counterpart for this source was detected in either
the \Chandra\ or \XMM\ images.

Using the \Chandra\ image alone, due to the smaller PSF, the morphology of the X-ray emission was analysed quantitatively. A
two dimensional $\beta$-model was fit to the X-ray image (including
instrumental effects) and the ellipticity of the best-fitting model was
$e=0.28\pm0.05$. The centroid shift (the standard deviation of the
separation between X-ray peak and centroid) was measured to be $\langle w
\rangle = (0.012\pm0.001)\rf$ (\rf\ is the radius corresponding to
$\Delta=500$ as computed in \textsection \ref{sec:temp-mass-prof}).  These
centroid shift and ellipticity values are close to the median values found
for a sample of 115 galaxy clusters of a range of redshifts observed with
\Chandra\ \citep{mau07b}. If only the more distant clusters ($z>0.5$) from
the \citet{mau07b} sample are considered, then the centroid shift of \jjj\
is lower than $21/32$ of those clusters. \jjj\ can thus be considered a
relatively relaxed example of a distant cluster.

In order to determine the mass of gas in the cluster, a surface
brightness profile was extracted from each image and converted into an
emission measure profile \citep[as described in][]{mau07a}. This
conversion assumed a metal abundance of $0.3\Zsol$, and a temperature
that varied with projected radius according to the mean temperature
profile found by \citet{vik06a}, normalised to the global temperature
and \rf\ measured for \jjj\ (\textsection
\ref{sec:glob-spectr-prop}). The conversion is only weakly dependent
on temperature so the choice of temperature profile does not
significantly affect our results. 

\begin{figure}
\begin{center}
\scalebox{0.33}{\includegraphics*[angle=270]{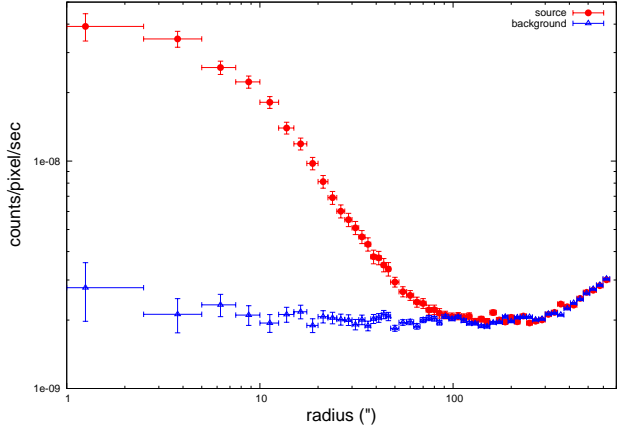}} \\
\caption{\label{f.sb} \XMM\ surface brightness profiles of \jjj\ and the
blank-sky background. The profiles were produced in the $0.3-2\keV$
band and the background profile was normalised to match the source
profile in regions free of source emission.}
\end{center}
\end{figure}

The background level in each bin was estimated from images produced
from blank-sky background files that were normalised to match the
count rates in the imaging energy band in regions free of source
emission. This was an iterative process, with the extent of the
cluster emission being determined from the profiles and then the
background images renormalised excluding this emission, with the
process repeated until convergence. The background normalisation is
illustrated in Fig. \ref{f.sb}, which shows the radial surface
brightness profiles of the source and background (after vignetting
correction) for the combined \XMM\ images. Vignetting correction of
the profiles was performed by dividing the counts in each bin by the
sum of the exposure map counts in that bin. Note the upward curve of
the profiles at large radii is due to the boosting of non-vignetted
particle events by the vignetting correction. This background component
is correctly subtracted, however, because the particle events are
boosted in the same way for the source and background images, and
level of particle events is similar in the source and background
datasets (\textsection
\ref{sec:backgr-prep}). The good agreement of the source and
background profiles outside of the cluster emission indicate that the
blank-sky background files provide an accurate background level for
the emissivity profiles.

A three-dimensional (3D) model emission measure profile was then
projected along the line of sight and fit simultaneously to the
observed \Chandra\ and
\XMM\ profiles. For the \XMM\ data, the projected profile was
convolved with a model of the \XMM\ PSF before the fit statistic was
computed. The PSF model used was a King
function\footnote{\url{http://xmm.vilspa.esa.es/docs/documents/CAL-TN-0018.pdf}}
with a core radius of $5.4\arcs$ and a slope of $1.5$. These
parameters are appropriate for $1.5\keV$ on-axis photons and are the
mean of the MOS and pn values.

\begin{figure}
\begin{center}
\scalebox{0.33}{\includegraphics*[angle=270]{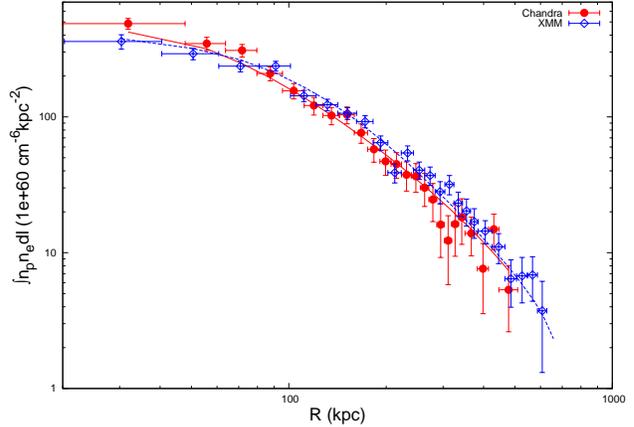}} \\
\caption{\label{f.em} The emission measure profile of \jjj\ measured
from the \Chandra\ and \XMM\ data are plotted along with the best
fitting model. The lines show the same 3D model projected along the
line of sight, and (in the case of \XMM) convolved with the telescope PSF.}
\end{center}
\end{figure}

The emission measure model used was the modified $\beta$-profile of
\citet{vik06a}:
\begin{eqnarray}\label{e.emis}
n_pn_e & = & n_0^2 \frac{(r/r_c)^{-\alpha}}{(1+r^2/r_c^2)^{3\beta-\alpha/2}}(1+r^\gamma/r_s^\gamma)^{-\epsilon/\gamma}.
\end{eqnarray}
The best-fitting model is plotted in Fig. \ref{f.em} along with the
observed profiles. The model was found to be a good fit to the
combined data, with $\chisq/\nu=45.8/43$ demonstrating good agreement
between the \Chandra\ and \XMM\ data. The best model
parameters were $n_0=1.31\times10^{-2}\pcc$ $r_c=55.7\kpc$,
$\alpha=0.0$, $\beta=0.46$, $r_s=872\kpc$, and $\epsilon=5.0$ with
$\gamma$ fixed at 3. Errors were not computed on individual parameters
as they are highly correlated, and model fits to Monte Carlo
randomisations of the data were used to determine the uncertainties on
the gas mass and other derived properties. The gas density is related
to the emission measure by $\rho_g=1.252m_p(n_pn_e)^{1/2}$, assuming a
cosmic plasma with helium abundances given by \citet{and89}.

While the morphology of \jjj\ is elliptical, our analysis assumes spherical
symmetry. However, this assumption leads to negligible errors ($\la3\%$) in
the total and gas mass compared to that derived using a triaxial model
\citep{pif03}. To allow comparisons with other clusters where such detailed
modeling of the gas distribution is not possible, the \Chandra\ surface
brightness profile was fit with a standard $\beta$-model \citep{cav76}. The
best-fitting model had a core radius of $r_c=84\pm5\kpc$ and a slope of
$\beta=0.60\pm0.02$.

\section{Spectral Analysis}
The general procedure followed for the analysis of the X-ray spectra of
\jjj\ was to extract spectra in a series of annular bins, and fit these to
create a temperature profile. The temperature profile was then used to
determine the mass profile of the cluster and define \rf. Spectra were then extracted
from within this radius and fit to give the global spectral properties of
the system.  All spectra were modeled by an absorbed APEC model
\citep{smi01}, with the absorbing column fixed at the Galactic value
\citep[$2.58\times10^{20}\pcmsq$; ][]{dic90}. The model temperature, metal abundance and normalisation
were free parameters in the fits, and the source spectra were grouped to
contain $\ge30$ counts per bin and fit using $\chisq$ minimisation. For
\XMM\ the pn and MOS spectra were fit simultaneously with the same model,
and similarly the data from the six
\Chandra\ observations were fit simultaneously for each region considered.

For both \Chandra\ and \XMM, background spectra were extracted from
the blank-sky background files using same detector regions as the source
spectra. The background spectra were normalised to match the high
energy count rate in the source datasets as discussed in \textsection
\ref{sec:backgr-prep}. In order to correct for the decrement in soft
Galactic foreground emission towards \jjj\ compared to the blank-sky
data, soft residual spectra were created by subtracting the blank-sky
spectrum from the local background spectrum in the same, source-free detector
region. In the case of \Chandra, the residual spectra were fit with a
thermal model with negative normalisation. This was then
included as a fixed component (scaled by extraction area) in
subsequent fits to the source spectra
\citep[this method is described in more detail in][]{vik05a}. For
\XMM, the soft residuals were scaled by extraction area and added onto
the background spectra \citep[a process known as double subtraction;
see e.g.][]{arn02b}. The \XMM\ Science Analysis System task {\it
evigweight} was applied to all of the \XMM\ source and background events lists
in order to account for differences in the telescope effective area
between the source region and the region from which the soft residuals
were derived. The entire analysis was also performed using a simple
local background spectrum extracted from the same dataset as the
source spectra, and no statistically significant differences were found in any of
the derived cluster properties.

\subsection{Temperature and Mass Profiles}\label{sec:temp-mass-prof}
\begin{figure}
\begin{center}
\scalebox{0.33}{\includegraphics*[angle=270]{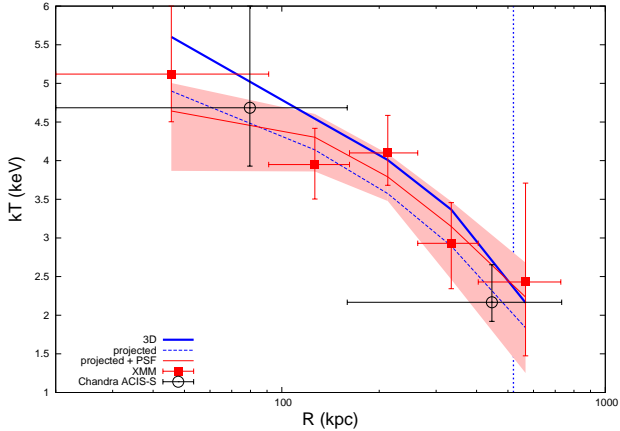}} \\
\caption{\label{f.kt} Projected temperature profiles measured for \jjj\
using \XMM\ and \Chandra. The curves show the best-fitting 3D model,
its projection, and the final model after PSF convolution, with the $1\sigma$ errors on the projected model shown by
the shaded region. The model was fit to the \XMM\ data alone, and the vertical line indicates \rf.}
\end{center}
\end{figure}

Spectra were extracted from annular regions centred on the cluster
centroid defined so that the combined background-subtracted counts
from the \XMM\ pn and MOS detectors were $\ge500$ in each region. The
same criterion was used to define regions for the \Chandra\ data. The
spectra were fit and the best-fitting temperatures are plotted in
Fig. \ref{f.kt}. The temperature profile was then fit with the
following model;
\begin{eqnarray}
kT(r) & = & kT_0\frac{(r/r_t)^{-a}}{(1+(r/r_t)^b)^{c/b}}.
\end{eqnarray}

This is the same model used by \citet{vik06a} to fit high-quality
temperature profiles of local relaxed clusters but the cool core
component has been removed from the model as the data for \jjj\ do not
require it. The model is thus a broken power law with a transition
region. The model was further simplified by setting $b=c/0.45$,
thereby fixing the width of the transition region to match that found
in the average fit to the \citet{vik06a} clusters. This 3D model was
projected along the line of sight to predict the temperature in each
bin of the observed profile. The projection weighted the different
temperature components using the measured gas density profile
according to the algorithm presented by \citet{vik06b}. Due to the
large radial bins used for the \Chandra\ data, and to simplify the
projection calculation, only the \XMM\ data were used for fitting the
temperature profile. 

The \XMM\ PSF is not negligible compared to the size of the radial
bins used for the temperature profile. The effect of the PSF is to
redistribute some photons that were emitted in one projected annulus
into a different projected annulus. The magnitude of this effect was
computed by using a background subtracted \Chandra\ image of
\jjj\ to provide the projected photon distribution with no PSF
redistribution. For each annular region in the temperature profile, a
sub-image was extracted, containing just the photons in that
region. This sub-image was then convolved with an image of the \XMM\
PSF (a PSF suitable for $1.5\keV$ photons detected on-axis was used),
and the number of photons redistributed from that annulus into  each
other annulus was measured. This process gave a redistribution matrix
describing the relative contribution of each projected annular region
to each bin in the final, projected, PSF-convolved profile. In the
fitting procedure, the temperature in each bin in the final model profile
was calculated by a second application of the \citet{vik06b} algorithm
to combine the contributions from every projected bin, weighted by the
PSF redistribution factors.

The best-fitting 3D model, its projection, and the final model including
PSF convolution are plotted in
Fig. \ref{f.kt}. The best-fitting model had parameters $kT_0=3.5\keV$,
$r_t=586\kpc$, $a=0.20$, and $c=1.55$. The probability distributions
of the parameters were determined by fits to Monte-Carlo
randomisations of the data, and used to determine uncertainties on
derived cluster properties. The best-fitting 3D model was also
compared with the \Chandra\ data plotted in Fig. \ref{f.kt}. The
projected temperature predicted by the 3D model for the inner and
outer \Chandra\ temperature bins are $4.6\keV$ and $2.8\keV$. The
prediction for inner bin agrees well with the measured value, while
that for the outer bin is slightly (but not significantly) higher than
the observed value ($2.2^{+0.5}_{-0.2}\keV$).

\begin{figure}
\begin{center}
\scalebox{0.33}{\includegraphics*[angle=270]{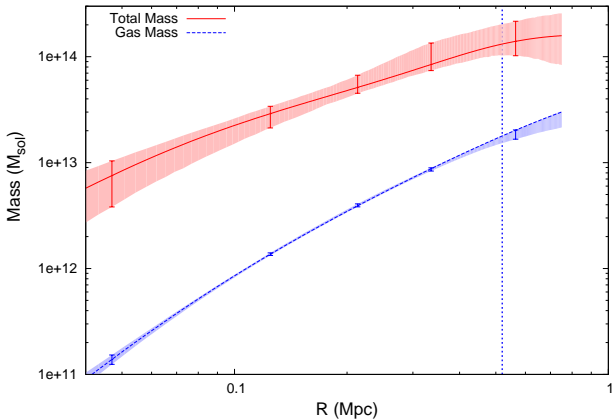}} \\
\caption{\label{f.mass} Profiles of the derived total and gas mass in
\jjj. The $1\sigma$ errors on the models are shown by the shaded
region, the data points mark the midpoints of the \XMM\ temperature profile
bins, and the vertical line indicates \rf.}  
\end{center}
\end{figure}

\begin{figure}
\begin{center}
\scalebox{0.33}{\includegraphics*[angle=270]{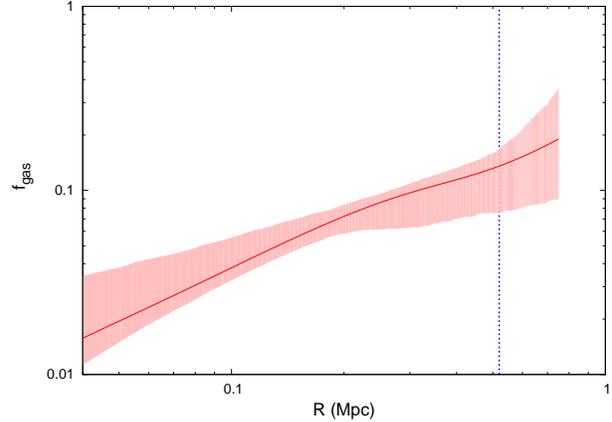}} \\
\caption{\label{f.fgas} Profiles of the gas mass fraction in
\jjj. The $1\sigma$ errors are shown by the shaded
region and the vertical line indicates \rf.}  
\end{center}
\end{figure}

The best-fitting 3D models for the gas density and temperature
profiles were then used to compute the total mass profile of \jjj\
under the assumption of hydrostatic equilibrium, propagating the Monte
Carlo uncertainties to all derived properties. This profile was then
used to compute the overdensity profile of the system and determine
$\rf=0.52^{+0.10}_{-0.05}\Mpc$ and
$\rn{2500}=0.23^{+0.06}_{-0.02}\Mpc$. The derived mass profile of
\jjj\ is plotted in Fig. \ref{f.mass}, along with the gas mass
profile.  The total mass of the system within \rf\ was found to be
$1.3^{+0.9}_{-0.3}\times10^{14}\Msol$. The gas mass fraction (\fgas)
profile of \jjj\ is plotted in Fig. \ref{f.fgas}. At \rf\ the gas mass
fraction was $0.14^{+0.02}_{-0.05}$, and within the smaller radius of
$\rn{2500}$, the gas fraction was $0.08^{+0.01}_{-0.02}$.

\subsection{Global Spectral Properties}\label{sec:glob-spectr-prop}
With \rf\ determined from the mass analysis, the integrated spectral
properties of \jjj\ were then measured within that aperture.  The central
$0.15\rf$ was excluded, as the emission from this region can be affected by
strong radiative cooling and merger events, leading to enhanced scatter
from the self-similar cluster scaling relations \citep[\egc][]{mau07b}. The
\XMM\ and \Chandra\ data were fit separately. The emission-weighted
temperature of the gas within \rf\ was found to be $3.4^{+0.3}_{-0.2}\keV$
based on the combined \XMM\ data, and $4.3^{+1.1}_{-0.7}\keV$ from the
combined \Chandra\ observations. The metallicity \citep[iron abundances
relative to the solar values of][]{and89} of the gas was measured to be
$0.18^{+0.17}_{-0.15}\Zsol$ from the \XMM\ data. While the data did not
permit a metallicity profile of the cluster, the value measured with the
core regions included increased slightly to $0.40^{+0.17}_{-0.16}\Zsol$,
perhaps indicative of an abundance peak in the core. The \Chandra\ data
were not able to provide useful constraints on the metallicity.

The bolometric X-ray luminosity of the cluster within \rf\ was measured
using the data from each satellite. The \XMM\ data gave
$\Lx=(3.4\pm0.4)\times10^{44}\ergps$
($\Lx=(4.6\pm0.4)\times10^{44}\ergps$ with the central $0.15\rf$ included),
while the \Chandra\ data gave $\Lx=3.0^{+1.0}_{-0.6}\times10^{44}\ergps$
($\Lx=4.5^{+0.9}_{-0.4}\times10^{44}\ergps$ including the
core). Uncertainties on the luminosities include those on the parameters of
the best-fit spectral models. The value of \Lx\ given above for the
\XMM\ data with the central $0.15\rf$ excluded includes a correction
for the redistribution of photons into and out of the excluded region by
the \XMM\ PSF. The same method used to calculated the redistribution
factors for the temperature profile was used to calculate that the PSF
redistribution leads to a net increase of $9\%$ in the number of
photons detected in the $(0.15-1)\rf$ aperture. The measured
luminosity was thus scaled down by this factor. 

The global properties of \jjj\ are all consistent with those measured
by \citet{pac07} using the original $20\ks$ of survey \XMM\ data.

\section{Scaling Relation Evolution}
The hydrostatic mass estimate obtained for \jjj\ presents a unique
opportunity to investigate the mass-observable scaling relations at
$z>1$. Recall that the gas mass and total mass were measured using a
combination of the \XMM\ and \Chandra\ data, although the total mass
estimate was strongly dependent on the \XMM\ temperature profile. The
global spectral properties were measured separately for \XMM\
and \Chandra, so datapoints for each satellite are included in the scaling
relations that follow. In all cases, the data were scaled by
an appropriate power of $E(z)$ (where $E^2(z) = \OM(1+z)^3 + (1-\OM-\Omega_\Lambda)(1+z)^2
+ \Omega_\Lambda$, describing the redshift evolution of the Hubble parameter) to remove the predicted self-similar evolution in each
relation. The exponent used for each relation is indicated in the
figures below.

\begin{figure}
\begin{center}
\scalebox{0.33}{\includegraphics*[angle=270]{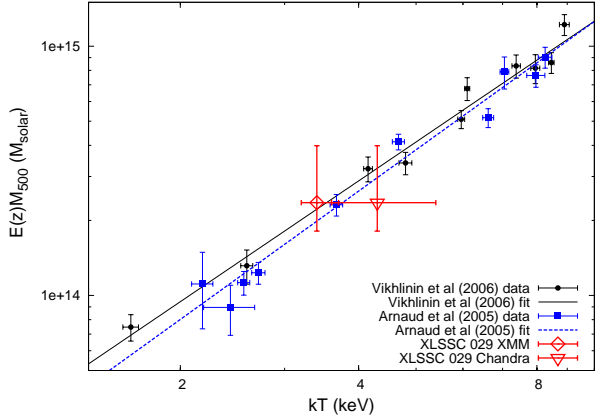}} \\
\caption{\label{f.mt} \jjj\ is plotted on the low-redshift mass
temperature relations of \citet{vik06a} and \citet{arn05}. The two \jjj\ points reflect the slightly different global temperatures measured with the \XMM\ and \Chandra\ data. The central $0.15\rf$ was excluded from the temperature measurements.}
\end{center}
\end{figure}

The mass and temperature of \jjj\ were compared with the data of
\citet{vik06a}, derived from high quality \Chandra\ observations of
low-redshift, relaxed clusters and of \citet{arn05}, derived from a
similar \XMM\ sample. The properties of \jjj\ were measured
in a consistent manner to those of these local samples. The
two datapoints for \jjj\ (for the \XMM\ and \Chandra\ temperatures)
are plotted along with the local data in
Fig. \ref{f.mt}. Both \jjj\ datapoints are consistent with the local
relations. Our data are also consistent with the mass-temperature
relation measured by \citep{kot05} from \XMM\ observations of clusters
at $0.4<z<0.7$.

\begin{figure}
\begin{center}
\scalebox{0.33}{\includegraphics*[angle=270]{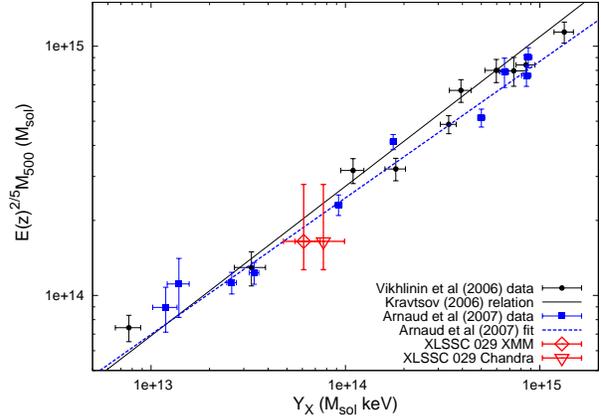}} \\
\caption{\label{f.ym} \jjj\ is plotted on the low-redshift \YM\ relations of \citet{vik06a} and \citet{arn05}. The two \jjj\ points reflect the slightly different global temperatures measured with the \XMM\ and \Chandra\ data. The central $0.15\rf$ was excluded from the temperature measurements.}
\end{center}
\end{figure}

Next, the \YM\ relation was investigated. \Yx, the product of the gas
mass and temperature, has been shown to have a low-scatter scaling
relation with cluster mass in simulated clusters
\citep{kra06a,poo07}. \Yx\ was calculated for \jjj, and was found to
be $(6.1\pm1.3)\times10^{13}\Msol\keV$ (using the \XMM\ kT; the
\Chandra\ kT gave $(7.7\pm2.2)\times10^{13}\Msol\keV$). These values
are plotted on the local \YM\ relations of \citet{vik06a} and
\citet[][]{arn07} in Fig. \ref{f.ym}. Again, the properties of \jjj\
are consistent with the local relation.

\begin{figure}
\begin{center}
\scalebox{0.33}{\includegraphics*[angle=270]{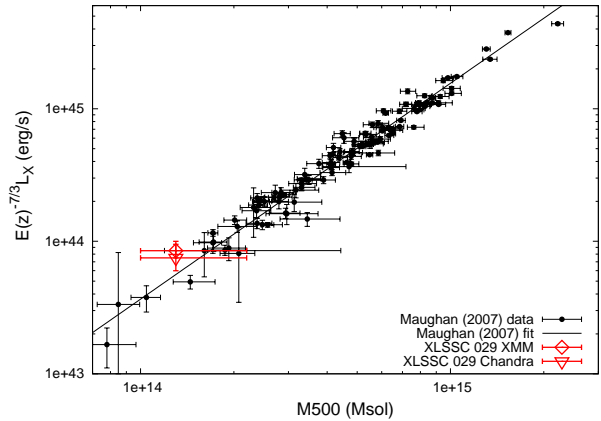}} \\
\caption{\label{f.lm} \jjj\ is plotted in the \LM\ relation of \citep{mau07b}. The two \jjj\ points reflect the slightly different luminosities measured with the \XMM\ and \Chandra\ data. The central $0.15\rf$ was excluded from the luminosity measurements.}
\end{center}
\end{figure}

Finally, in Fig. \ref{f.lm} we plot \jjj\ on the \LM\ relation of \citet{mau07b}. This is a sample of 115 clusters over the redshift range $0.1<z<1.3$ observed with \Chandra. For those clusters, the masses were estimated from their \Yx\ values. For \jjj, however, we have the advantage that the mass was estimated from a full hydrostatic mass analysis. Once more, the properties of \jjj\ are consistent with the self-similar evolution of the cluster population.

\section{Discussion}
The deep \XMM\ and \Chandra\ observations of \jjj\ have enabled a uniquely
detailed study of a $z>1$ galaxy cluster. This allows the comparison
of the properties of \jjj\ with the expected evolution of the local cluster
population. The iron abundance in \jjj\ is not tightly constrained by the
data, but the value measured with the central $0.15\rf$ excluded is
consistent with the observed and predicted evolution in ICM metal abundance
\citep{ett05,bal07,mau08a}. The increased metal abundance when the core
regions are included, while not statistically significant, is also in line
with the mean trends found by \citet{mau08a}.

A key result from this analysis is the measurement of the temperature
profile of \jjj, providing the foundation for the hydrostatic mass analysis
of the system. The accuracy of the \XMM\ temperature profile is supported
by the agreement of the cruder \Chandra\ profile. The modeling of the
temperature profile is a key step in the mass analysis, and we note that
while the model used provides a good description of the radial temperature
distribution in \jjj, other models could also be used to fit the
data. This is not a concern, as we do not require a physical interpretation of the
individual model parameters; the model is simply used to give the value
and gradient of the 3D temperature. Importantly we do not extrapolate our
mass estimates beyond the range of the data, so the results are not
strongly dependent on the choice of temperature profile model. That said,
the model values from beyond the projected radial range of the data do have
some effect on the fit, because gas from all external radii is projected
along the line of sight in the fitting process. This effect is negligible
because the low density of the gas at large radii minimises its contribution to
the projected temperature at the projected radii of interest.

As discussed in \textsection\ref{s.img} the analysis of the X-ray images
suggests that \jjj\ is a relatively relaxed cluster, particularly compared
to the high-redshift cluster population. The X-ray mass estimate requires
that the ICM is in hydrostatic equilibrium with the gravitational
potential. While the X-ray morphology suggests that this is the case, there
are examples of clusters with relaxed X-ray morphologies that show evidence
for being out of hydrostatic equilibrium \citep[e.g. CLJ1226.9+3332 at
z=0.89;][]{mau07a}. In the case of CLJ1226.9+3332, a temperature map showed
asymmetric structure coincident with a subclump of galaxies detected in the
optical images. The X-ray data are insufficient for a spectral map of \jjj\
(to achieve a similar quality map to that of CLJ1226.9+3332 would require
almost ten times the current \XMM\ exposure), but the available optical
data do not indicate substructure in the galaxy distribution.

\begin{figure}
\begin{center}
\scalebox{0.33}{\includegraphics*[angle=270]{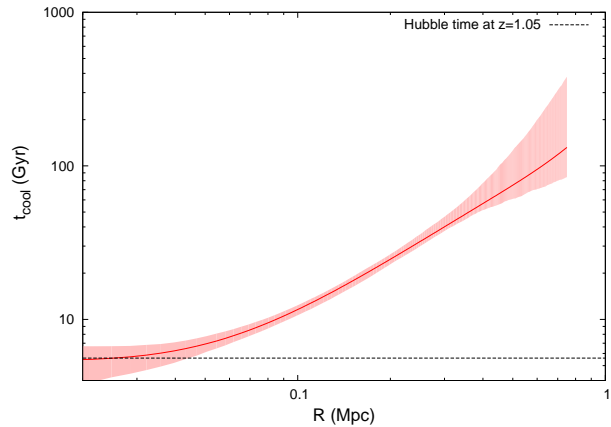}} \\
\caption{\label{f.tc} Cooling time profile of \jjj. The shaded region shows
the $1\sigma$ uncertainties, and the dashed line indicates the Hubble time
at the redshift of \jjj.}
\end{center}
\end{figure}

Many relaxed clusters in the local universe exhibit cool cores due to the
short radiative cooling times in the central regions
\citep[\egc][]{fab94b}. The fraction of cool core clusters has been found
to be significantly lower at redshifts higher than $0.5$
\citep{vik07}. This is believed to be due to the higher incidence of
cluster mergers at high redshift, with mergers disrupting the cooling
process. The temperature profile of \jjj\ gives no indication of any cool
core in the system. This could indicate that the system has recently
undergone a merger event, or it could simply be because the gas in the
cluster core has not had long enough to cool significantly. In order to
investigate the latter possibility, the radiative cooling time of the ICM
was calculated using the measured gas density and temperature profile
according to
\begin{eqnarray}
t_{cool} & = &
8.5\times10^{10}\yr\left(\frac{n_p}{10^{-3}\pcc}\right)^{-1}\left(\frac{T}{10^{8}\K}\right)^{1/2}
\end{eqnarray}
\citep{sar86}, and the resulting profile is plotted in
Fig. \ref{f.tc}. The profile shows that the radiative cooling time of the
ICM in \jjj\ is longer than the Hubble time (at $z=1.05$) outside of the
central $\sim30\kpc$, so even without recent merger activity, there may not
have been sufficient time for a significant cool core to develop in this
system.

The gas mass fractions measured for \jjj\ at \rf\ and \rn{2500} are in good
agreement with the corresponding values measured for the \citet{vik06a} clusters. This
is consistent with no evolution in \fgas\ in our assumed \LCDM\
cosmology. With the reliable masses measured in these deep observations,
\jjj\ can provide a useful high-redshift datapoint for cosmological studies
that make use of \fgas\ evolution \citep{all04}. 


Perhaps the most interesting results from the study of \jjj\ are obtained
by its use as a probe of the evolution of the mass-observable scaling
relations. In all of the relations tested, the observable properties of
\jjj\ were consistent with the self-similar evolution of the scaling
relations. In particular, the agreement of \jjj\ with the
\YM\ relation provides further support for the use of \Yx\ as a mass proxy
for even the most distant clusters \citep[see
also][]{kra06a,mau07b}. Furthermore, \jjj\ was found to lie on the \LM\
relation of \citet{mau07b}. This provides additional support for their
conclusion that
\Lx\ can be used as an effective mass proxy, as the mass of \jjj\ was
estimated from a full hydrostatic analysis rather than from \Yx\ as in the
\citet{mau07b} sample.

A similar study was performed on the $z=0.89$ cluster CLJ1226.9+3332, but
that cluster was found to deviate from the self-similar scaling relations
\citep{mau07a}. This was deemed likely to be due to merger activity in the
that cluster. In the case of \jjj, the current data show no evidence for
merger activity, but are not sufficiently deep to detect temperature
substructure like that found in CLJ1226.9+3332. The relaxed appearance of
\jjj\ and its agreement with the self-similar evolution of mass-observable
relations form a self-consistent picture. However it is possible that
biases on our mass estimate due to a deviation from hydrostatic equilibrium
and non-standard evolution could conspire to cancel one-another out.

\section{Conclusions}
Our hydrostatic X-ray mass analysis of \jjj\ has provided a unique
opportunity to test the evolution of the mass-observable scaling relations
at $z=1$. While the strength of our conclusions are limited by the use of a
single object, we found no evidence to reject the simple self-similar
evolution of the scaling relations. This provides support for the use of
these scaling relations in cosmological studies with large sample of
distant clusters where the data do not allow such detailed mass analyses.

\section{Acknowledgments}
We are grateful to the referee for several useful suggestions,
particularly on improving the treatment of the \XMM\ PSF. BJM was
supported during most of this work by NASA through Chandra
Postdoctoral Fellowship Award Number PF4-50034 issued by the Chandra
X-ray Observatory Center, which is operated by the Smithsonian
Astrophysical Observatory for and on behalf of NASA under contract
NAS8-03060. SA acknowledges financial support from contract ASI-INAF
I/023/05/0.

\bibliographystyle{mn2e}
\bibliography{clusters}

\end{document}